\documentclass[letterpaper,twocolumn,superscriptaddress,floatfix]{revtex4}
\usepackage{inputenc}
\usepackage{bm}
\usepackage{multirow,amssymb,amsbsy,amsmath}
\usepackage{graphicx}
\usepackage{float}
\usepackage{verbatim}
\makeatletter
\usepackage{pifont}
\usepackage{upgreek}
\usepackage{color}
\usepackage{indentfirst}
\usepackage{soul}
\usepackage{caption}
\usepackage{dcolumn}
\usepackage{ifthen}
\usepackage{booktabs}
\usepackage{nicefrac}

\usepackage{dsfont}

\makeatother
\makeatletter 
\renewcommand{\section}{\@startsection{section}{1}{0mm}
  {-\baselineskip}{0.5\baselineskip}{\bf\leftline}}
\makeatother

\begin{document}
\title{Laser Direct Writing of Visible Spin Defects in Hexagonal Boron Nitride for Applications in Spin-Based Technologies}

\affiliation{CAS Key Laboratory of Quantum Information, University of Science and Technology of China, Hefei, Anhui 230026, China}
\affiliation{CAS Center For Excellence in Quantum Information and Quantum Physics,
University of Science and Technology of China, Hefei, Anhui 230026, China}
\affiliation{School of Physics and Materials Engineering, Hefei Normal University, Hefei, Anhui 230601, China}
\affiliation{Hefei National Laboratory, University of Science and Technology of China, Hefei, Anhui 230088, China}

\author{Yuan-Ze Yang}
\thanks{These authors contribute equally to this work.}
\affiliation{CAS Key Laboratory of Quantum Information, University of Science and Technology of China, Hefei, Anhui 230026, China}
\affiliation{CAS Center For Excellence in Quantum Information and Quantum Physics,
University of Science and Technology of China, Hefei, Anhui 230026, China}
\affiliation{Hefei National Laboratory, University of Science and Technology of China, Hefei, Anhui 230088, China}
\author{Tian-Xiang Zhu}
\thanks{These authors contribute equally to this work.}
\affiliation{CAS Key Laboratory of Quantum Information, University of Science and Technology of China, Hefei, Anhui 230026, China}
\affiliation{CAS Center For Excellence in Quantum Information and Quantum Physics,
University of Science and Technology of China, Hefei, Anhui 230026, China}
\affiliation{Hefei National Laboratory, University of Science and Technology of China, Hefei, Anhui 230088, China}
\author{Zhi-Peng Li}
\affiliation{CAS Key Laboratory of Quantum Information, University of Science and Technology of China, Hefei, Anhui 230026, China}
\affiliation{CAS Center For Excellence in Quantum Information and Quantum Physics,
University of Science and Technology of China, Hefei, Anhui 230026, China}
\affiliation{Hefei National Laboratory, University of Science and Technology of China, Hefei, Anhui 230088, China}
\author{Xiao-Dong Zeng}
\affiliation{CAS Key Laboratory of Quantum Information, University of Science and Technology of China, Hefei, Anhui 230026, China}
\affiliation{CAS Center For Excellence in Quantum Information and Quantum Physics,
University of Science and Technology of China, Hefei, Anhui 230026, China}
\affiliation{Hefei National Laboratory, University of Science and Technology of China, Hefei, Anhui 230088, China}
\author{Nai-Jie Guo}
\affiliation{CAS Key Laboratory of Quantum Information, University of Science and Technology of China, Hefei, Anhui 230026, China}
\affiliation{CAS Center For Excellence in Quantum Information and Quantum Physics,
University of Science and Technology of China, Hefei, Anhui 230026, China}
\affiliation{Hefei National Laboratory, University of Science and Technology of China, Hefei, Anhui 230088, China}
\author{Shang Yu}
\affiliation{CAS Key Laboratory of Quantum Information, University of Science and Technology of China, Hefei, Anhui 230026, China}
\affiliation{CAS Center For Excellence in Quantum Information and Quantum Physics,
University of Science and Technology of China, Hefei, Anhui 230026, China}
\affiliation{Hefei National Laboratory, University of Science and Technology of China, Hefei, Anhui 230088, China}
\author{Yu Meng}
\affiliation{CAS Key Laboratory of Quantum Information, University of Science and Technology of China, Hefei, Anhui 230026, China}
\affiliation{CAS Center For Excellence in Quantum Information and Quantum Physics,
University of Science and Technology of China, Hefei, Anhui 230026, China}
\affiliation{Hefei National Laboratory, University of Science and Technology of China, Hefei, Anhui 230088, China}
\author{Zhao-An Wang}
\affiliation{CAS Key Laboratory of Quantum Information, University of Science and Technology of China, Hefei, Anhui 230026, China}
\affiliation{CAS Center For Excellence in Quantum Information and Quantum Physics,
University of Science and Technology of China, Hefei, Anhui 230026, China}
\affiliation{Hefei National Laboratory, University of Science and Technology of China, Hefei, Anhui 230088, China}
\author{Lin-Ke Xie}
\affiliation{CAS Key Laboratory of Quantum Information, University of Science and Technology of China, Hefei, Anhui 230026, China}
\affiliation{CAS Center For Excellence in Quantum Information and Quantum Physics,
University of Science and Technology of China, Hefei, Anhui 230026, China}
\affiliation{Hefei National Laboratory, University of Science and Technology of China, Hefei, Anhui 230088, China}
\author{Zong-Quan Zhou}
\affiliation{CAS Key Laboratory of Quantum Information, University of Science and Technology of China, Hefei, Anhui 230026, China}
\affiliation{CAS Center For Excellence in Quantum Information and Quantum Physics,
University of Science and Technology of China, Hefei, Anhui 230026, China}
\affiliation{Hefei National Laboratory, University of Science and Technology of China, Hefei, Anhui 230088, China}
\author{Qiang Li}
\affiliation{CAS Key Laboratory of Quantum Information, University of Science and Technology of China, Hefei, Anhui 230026, China}
\affiliation{CAS Center For Excellence in Quantum Information and Quantum Physics,
University of Science and Technology of China, Hefei, Anhui 230026, China}
\affiliation{Hefei National Laboratory, University of Science and Technology of China, Hefei, Anhui 230088, China}
\author{Jin-Shi Xu}
\affiliation{CAS Key Laboratory of Quantum Information, University of Science and Technology of China, Hefei, Anhui 230026, China}
\affiliation{CAS Center For Excellence in Quantum Information and Quantum Physics,
University of Science and Technology of China, Hefei, Anhui 230026, China}
\affiliation{Hefei National Laboratory, University of Science and Technology of China, Hefei, Anhui 230088, China}
\author{Xiao-Ying Gao}
\affiliation{CAS Key Laboratory of Crust-Mantle Materials and Environments, School of Earth and Space Sciences, University of Science and Technology of China, Hefei, 230026, China}
\author{Wei Liu}
\altaffiliation{Email: lw691225@ustc.edu.cn}
\affiliation{CAS Key Laboratory of Quantum Information, University of Science and Technology of China, Hefei, Anhui 230026, China}
\affiliation{CAS Center For Excellence in Quantum Information and Quantum Physics,
University of Science and Technology of China, Hefei, Anhui 230026, China}
\affiliation{School of Physics and Materials Engineering, Hefei Normal University, Hefei, Anhui 230601, China}
\affiliation{Hefei National Laboratory, University of Science and Technology of China, Hefei, Anhui 230088, China}
\author{Yi-Tao Wang}
\altaffiliation{Email: yitao@ustc.edu.cn}
\affiliation{CAS Key Laboratory of Quantum Information, University of Science and Technology of China, Hefei, Anhui 230026, China}
\affiliation{CAS Center For Excellence in Quantum Information and Quantum Physics,
University of Science and Technology of China, Hefei, Anhui 230026, China}
\affiliation{Hefei National Laboratory, University of Science and Technology of China, Hefei, Anhui 230088, China}
\author{Jian-Shun Tang}
\altaffiliation{Email: tjs@ustc.edu.cn}
\affiliation{CAS Key Laboratory of Quantum Information, University of Science and Technology of China, Hefei, Anhui 230026, China}
\affiliation{CAS Center For Excellence in Quantum Information and Quantum Physics,
University of Science and Technology of China, Hefei, Anhui 230026, China}
\affiliation{Hefei National Laboratory, University of Science and Technology of China, Hefei, Anhui 230088, China}
\author{Chuan-Feng Li}
\altaffiliation{Email: cfli@ustc.edu.cn}
\affiliation{CAS Key Laboratory of Quantum Information, University of Science and Technology of China, Hefei, Anhui 230026, China}
\affiliation{CAS Center For Excellence in Quantum Information and Quantum Physics,
University of Science and Technology of China, Hefei, Anhui 230026, China}
\affiliation{Hefei National Laboratory, University of Science and Technology of China, Hefei, Anhui 230088, China}
\author{Guang-Can Guo}
\affiliation{CAS Key Laboratory of Quantum Information, University of Science and Technology of China, Hefei, Anhui 230026, China}
\affiliation{CAS Center For Excellence in Quantum Information and Quantum Physics,
University of Science and Technology of China, Hefei, Anhui 230026, China}
\affiliation{Hefei National Laboratory, University of Science and Technology of China, Hefei, Anhui 230088, China}

\begin{abstract}
Optically addressable spins in two-dimensional hexagonal boron nitride (hBN) attract widespread attention for their potential advantage in on-chip quantum devices, such as quantum sensors and quantum network. A variety of spin defects have been found in hBN, but no convenient and deterministic generation methods have been reported for other defects except negatively charged boron vacancy ($\rm V_B^-$). Here we report that by using femtosecond laser direct writing technology, we can deterministically create spin defect ensembles with spectra range from 550 nm to 800 nm on nanoscale hBN flakes. Positive single-peak optically detected magnetic resonance (ODMR) signals are detected in the presence of magnetic field perpendicular to the substrate, and the contrast can reach 0.8$\rm \%$. With the appropriate thickness of hBN flakes, substrate and femtosecond laser pulse energy, we can deterministically and efficiently generate bright spin defect array. Our results provide a convenient deterministic method to create spin defects in hBN, which will motivate more endeavors for future researches and applications of spin-based technologies such as quantum magnetometer array.
\end{abstract}

\maketitle
\date{\today}
\section*{Introduction}

Optically addressable spin defects in wide band-gap materials are available for the applications in quantum information \cite{QInfor1,QInfor2Error,QT,QRNV} and quantum sensing \cite{QSens1,QSens2,QSens3Yan,QSen4hBN,NVSky}. Besides solid-state spins in diamond \cite{NVFem, NVSky,NVSpinEntan,NVReview,NVEnhance} and silicon carbide \cite{SiCFem, SiCScience,SiCNmat,LiSiC,SiCControl,SiCSpintronics,SiCGene}, hexagonal boron nitride (hBN) -- a van der Waals material with wide bandgap ($\sim$ 6 eV) and excellent scalability to host on-chip systems \cite{Chip1,Chip2} -- have been reported to have optically active spin defects. Since Tran et \emph{al.} \cite{TTTnano} first discovered single photon emission in monolayer hBN at room temperature, research into quantum emitters in hBN began to flourish \cite{Stab1, Stab2, Femto2, TTTElecAnneal, SolventEx}. So far, there have been found several spin defects with optically detected magnetic resonance (ODMR) signal in hBN, indicating the great potential of hBN in quantum technology. However, for most spin defects, we have not yet mastered the technology for controllable fabrication.

Negatively charged boron vacancy ($\rm V_B^-$) defect is the most studied spin defect in hBN \cite{Electron, NeutronFirst, TemLiu, RabiLiu, Femto1, VB-Cal}. The structure and properties have been deeply researched in experiment and theory, and $\rm V_B^-$ ensemble can be easily fabricated by neutron irradiation \cite{NeutronFirst,TemLiu,RabiLiu}, electron irradiation \cite{Electron}, femtosecond laser ablation \cite{Femto1} and ion implantation \cite{FIB,IonGuo}, etc. Recently, given the two-dimensional (2D) nature of hBN, $\rm V_B^-$ has been demonstrated to show promising performance for in-situ quantum sensing \cite{hBNSenArray1, hBNSenArray2,hBNSenArray3}, and has become a research hotspot. Moreover, relatively a few studies have been carried out on other spin defects including defects related with carbon and some defects with zero phonon line (ZPL) at around 720 $\rm nm$ and 545 $\rm nm$ \cite{CIgor,CCambridge,CWTrup,GuoRabi}. These defects have many outstanding properties, such as high brightness, high Debye-Waller (DW) factor, and isolated single spins also have been found. Although the carbon related spin defect can be generated by metal organic vapour phase epitaxy (MOVPE) by controlling the gas environment \cite{CIgor,CCambridge}, but this method is a little bit complicated and the defects locate randomly; as for directly annealing, the generation of spin defect with ZPL $\sim$ 545 nm seems more uncontrollable \cite{GuoRabi}. For further research, a key step is to find a convenient deterministic fabrication method, which is also the basis for practical applications.

\begin{figure*}[ht]
    \centering
    \includegraphics[width=0.85\linewidth]{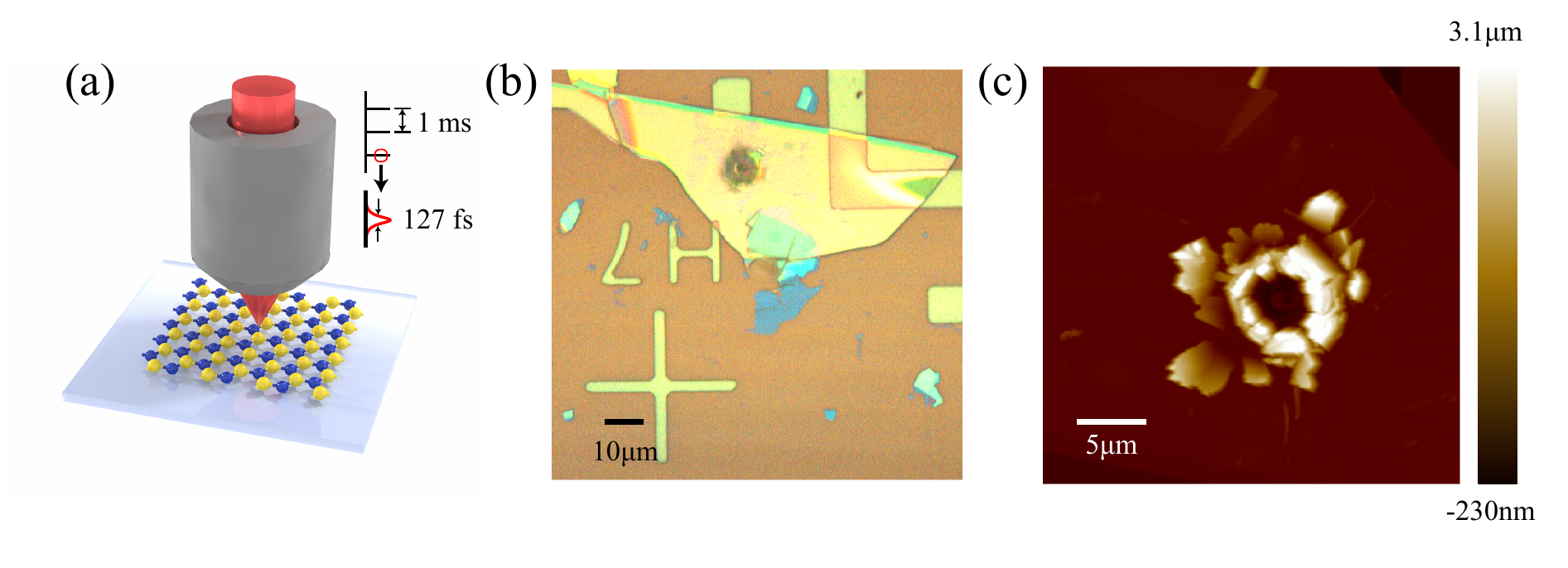}
    \caption{(a) Schematic of the femtosecond laser direct writing. High-energy femtosecond laser is focused on the hBN sample surface using a 100$\rm \times$ objective lens with N.A. = 0.7.  The repetition frequency of the femtosecond laser is set to 1 kHz, and the FWHM of a single pulse is 210 fs.
    (b) The optical microscopy image of a laser-irradiated hBN flakes. The femtosecond laser pulse number was 10 while the single pulse energy was 4.98 $\rm \upmu J$. A $\sim$ 5-$\rm \upmu m$ diameter hole appears where a femtosecond laser was focused. The scale bar is 10 $\rm \upmu m$.
    (c) Characterization results of the laser ablation region in (b) by atomic force microscope (AFM). The hBN flake was broken into irregular fragments with the average thickness more than 2 $\rm \upmu m$ while that of flat region is 255 $\rm nm$. A negative value for the height of the ablation spot indicates that the femtosecond laser has completely broken the hBN and also broken part of the substrate. The scale bar is 5 $\rm \upmu m$.}
    \label{Figure 1}
\end{figure*}

Here, we demonstrate a convenient deterministic method to generate spin defect ensemble in nanoscale hBN flakes by direct femtosecond laser writing, followed by vacuum annealing. The photoluminescence (PL) spectra of generated defect ensembles all exhibit a big packet in visible range among 550 nm and 800 nm at room temperature, which is different with other reported spin defects. The ODMR spectra show a positive peak without hyperfine structure even at low microwave power, and the contrast can achieve 0.8\%. In addition, we varied the thickness of hBN, the energy of femtosecond laser pulse, and the type of substrate to explore the best generation conditions of the spin defects. Under the appropriate conditions, we can deterministically and efficiently generate spin defect array with low damage to the sample, which can be used as quantum magnetic sensor array \cite{hBNSenArray1, hBNSenArray2, hBNSenArray3, sensor1, sensor2, sensor3, sensor4}. Our work provides a powerful method to generate spin defects in hBN, and paves a way for future studies and spin-based quantum technologies.

\section*{Results}
Figure 1 shows the schematic of our femtosecond laser micromachining (FLM) system and the characterization results after femtosecond laser irradiation. The repetition rate of the FLM system was set to 1 kHz with wavelength fixed at 1030 nm. Ten femtosecond laser pulses with full width of the half maximum (FWHM) of 210 fs and single pulse energy of 4.98 $\rm \upmu J$ were focused on the surface of hBN flake on Si/SiO$_2$ substrate by a 100$\rm\times$ objective (see Figure 1(a)). After annealing in vacuum environment in the tube furnace for 120 minutes at 1000 $\rm ^{\circ}$C, surface morphological characterization was carried out using optical microscope (Figure 1(b)) and atomic force microscope (Figure 1(c)), respectively. More details can be found in Section Methods.

After femtosecond laser writing, a $\sim$5-$\upmu$m hole was left on the flake, and some hBN fragments with triangular and quadrilateral petal-like appearance were found with thickness over 2 $\upmu$m around the hole of the sample. In this experiment, we used a homebuilt confocal microscopy system combined with a microwave system to characterize our sample at room temperature. As shown in Figure 2(a), a PL map of the same sample in Figure 1(b) is displayed. The PL of the sample around the ablation hole is significantly brighter than that of the sample in other areas, which indicates that the femtosecond laser irradiation is a necessary condition for the generation of these fluorescent defects. Figure 2(b) shows the count rate of the emitters in the black circle area of Figure 2(a) during 300 s with 1-mW excitation laser. This emission is bright (the count rate exceeding 1500 k/s) and stable. The normalized PL spectrum of the defect ensemble in the same area shows a large packet in visible light band, as shown in Figure 2(c). Two relatively dominant peaks are at around 620 nm and 720 nm, respectively. The Raman spectrum only shows Raman peaks of hBN ($\rm1366.8$ $\rm cm^{-1}$) and Si ($\rm520.7$ $\rm cm^{-1}$) (Figure S8(a) in Supporting Information). In addition, the Raman spectra of samples with spin defects generated on Au and quartz substrates both only show Raman peaks of hBN (Figure S8(b,h) in Supporting Information). These results indicate that the spin defects are embedded in hBN rather than other materials, such as cubic boron nitride (Raman peak around $\rm1295$ $\rm cm^{-1}$\cite{Femto2}).

\begin{figure}[htb]
    \centering
    \includegraphics[width=1\linewidth]{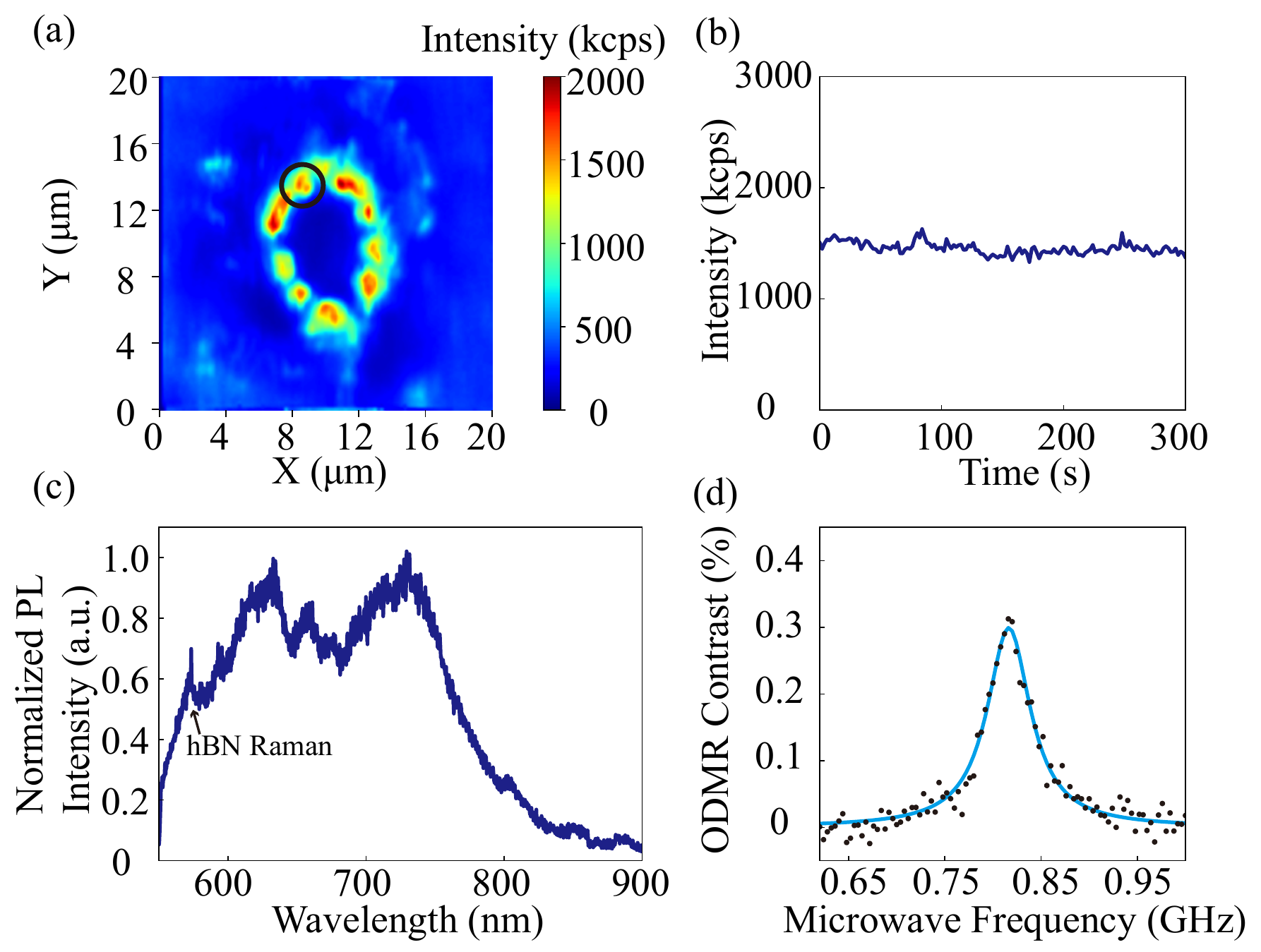}
    \caption{(a) The confocal PL image of the laser written hBN in Figure 1(b,c). Under the 532-nm excitation laser of 1 $\rm mW$, the region of hBN fragments around the ablation spot fluoresces brightly, most of which count over 1$\times 10^6$ per second. (b) The PL emission stability diagram of the emitter circled in (a). The emission intensity remains stable in 300 s. (c) Normalized PL spectrum of the same emitter. The shape of the spectrum is a large packet that covers the measurement range, and has two obvious peaks around 620 nm and 720 nm, respectively. (d) ODMR spectrum of the same emitter under magnetic field perpendicular to the substrate at 29.5 mT at room temperature. The solid line is the fitting result of Lorentzian function.}
    \label{Figure 2}
\end{figure}

Then, an ODMR measurement was carried out by sweeping the microwave (MW) frequency from 600 MHz to 1000 MHz, when a magnetic field perpendicular to the substrate with intensity of 29.5 mT was applied to the sample. A positive single-peak ODMR spectrum centered at $\sim$ 816 MHz was measured (Figure 2(d)). The fitting result with Lorentzian function (solid line) shows that the contrast is $0.30\%$ and the linewidth is 54.6 MHz. In addition, we need to point out that we did not observe such ODMR signals in the unannealed samples.

Next we focused on the spin properties of these defects. We measured the ODMR spectra of the spin defect ensemble (ensemble E1, see Supporting Information Figure S1) under different magnetic field perpendicular to the substrate, and five representative ODMR spectra with Lorentzian fits are shown in Figure 3(a). Figure 3(b) presents the magnetic-field-dependence of ODMR peak frequency, and a linear fit (solid line) reveals a g-factor of 1.92 $\pm$ 0.25, close to the value for the free electron. Further, the expected behavior of the ODMR contrast of defect ensemble (ensemble E2, see Supporting Information Figure S1) as a function of microwave power and excitation-laser power has been studied at 29.5 mT (Figure 3(c,d)). The fitting function (solid lines) is $C=(C_{sat} \times P)/(P + P_{sat})$. The saturation power ($P_{sat}^{MW}$) of the microwave power is 1.52 mW, with laser power fixed at 1 mW. Varying the excitation laser power from 0 to 1 mW, the ODMR contrast increases and gradually become saturated with microwave power fixed at 14.1 mW.

\begin{figure}[ht]
    \centering
    \includegraphics[width=1\linewidth]{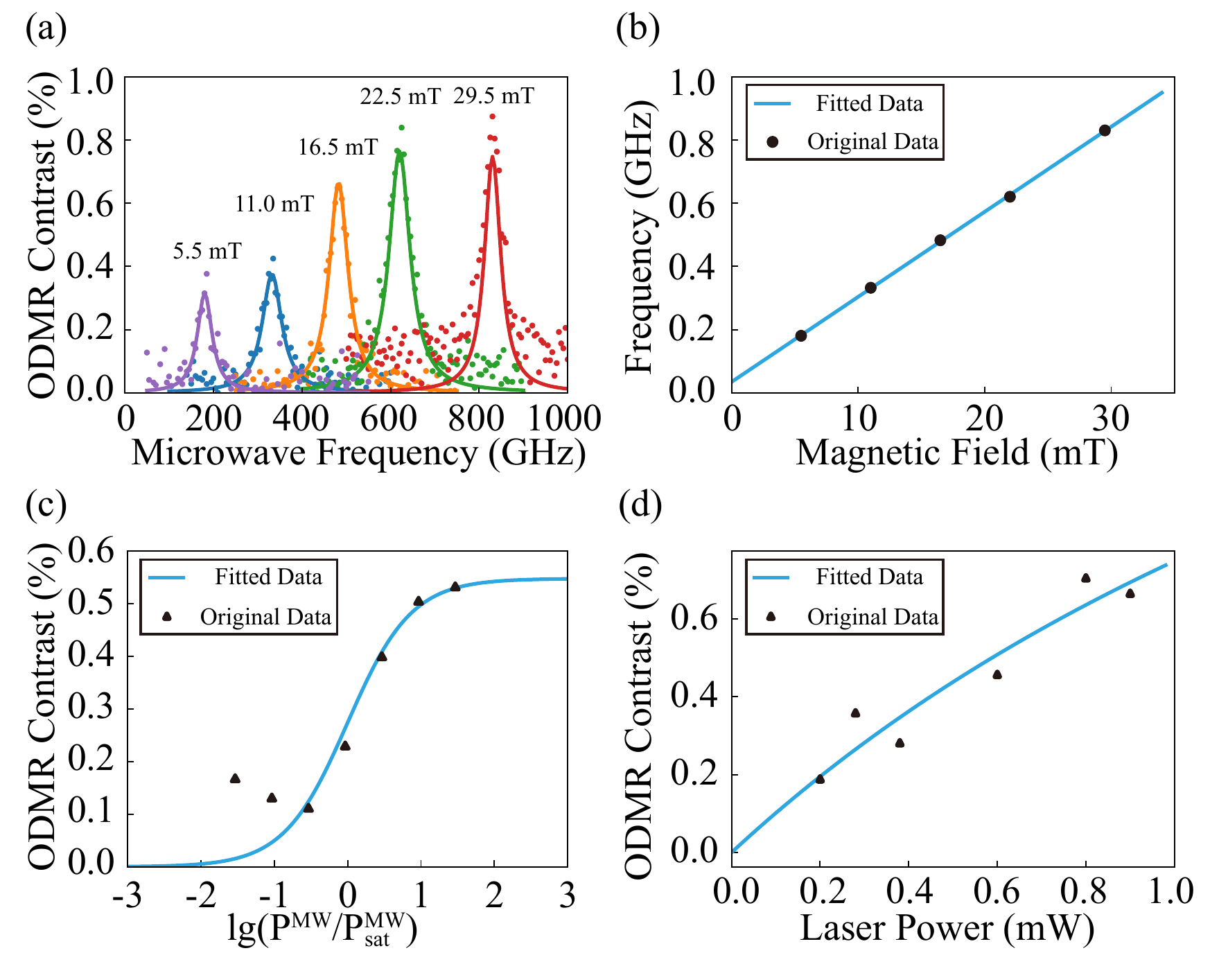}
    \caption{(a) ODMR spectra of an emitter under different external magnetic field perpendicular to the substrate with fixed pump power and microwave power. (b) ODMR central frequency as a function of the intensity of magnetic field parallel to the c-axis. The linear fitting result (solid line) reveals the g = 1.92 $\pm$ 0.25, corresponds to the value for the spin. (c) Dependence of the ODMR contrast on normalized microwave power ($P^{MW}/P_{sat}^{MW}$) measured at 1 mW excitation laser, where the saturation power $P_{sat}^{MW}$ is 1.52 mW. (d) ODMR contrast as a function of power of 532-$\rm nm$ laser at 14.1 mW microwave. The solid lines is the fitting result with $C=C_{sat} \times P/(P + P_{sat})$.}
    \label{Figure 3}
\end{figure}

\begin{figure*}[ht]
    \centering
    \includegraphics[width=1\linewidth]{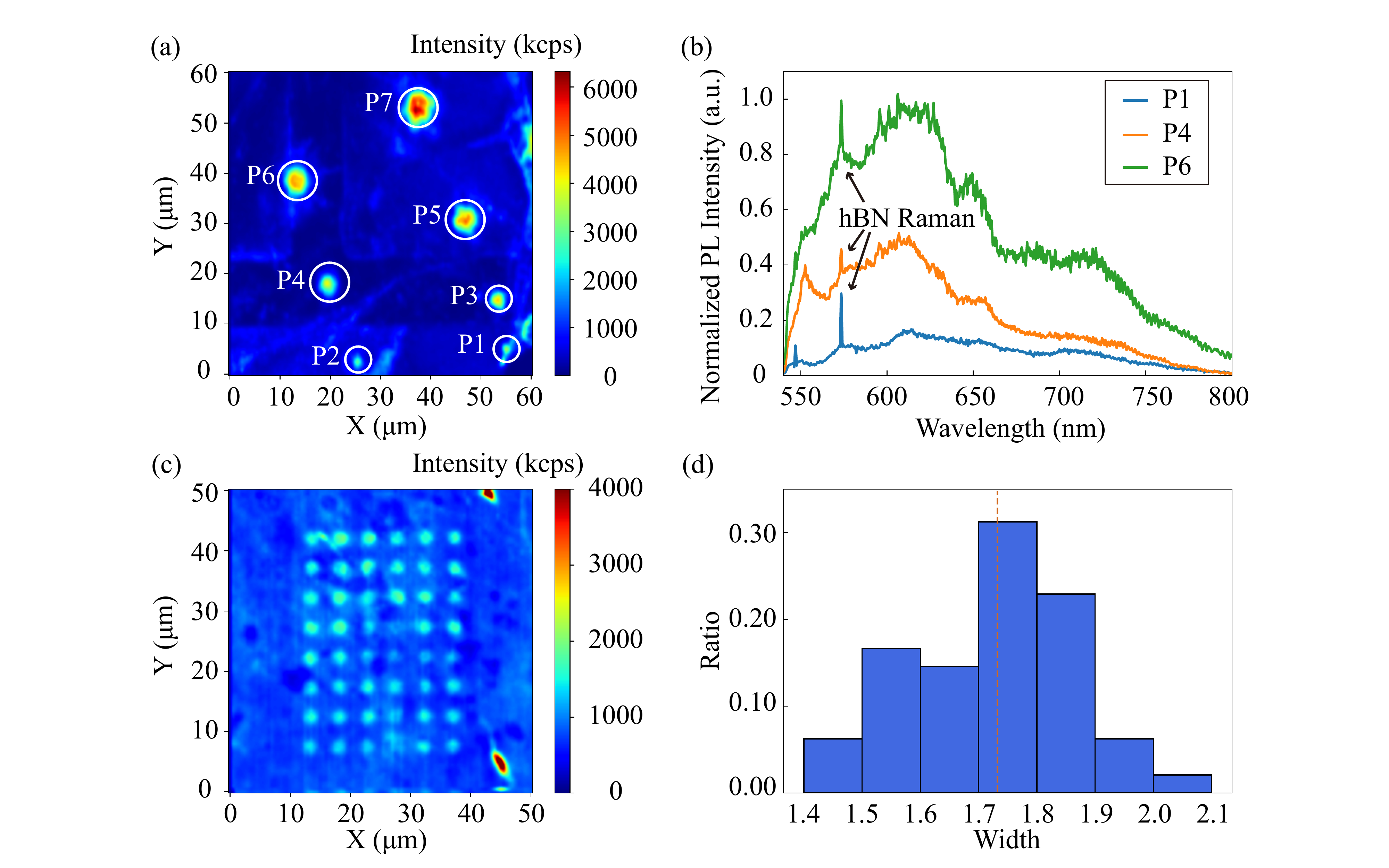}
    \caption{Results of different femtosecond laser energy and stable. (a) The PL image of a thick hBN flake irradiated with different laser energies. The single pulse energy of the femtosecond laser varies in 0.25 $\rm \upmu J$, 0.62 $\rm \upmu J$, 1.25 $\rm \upmu J$, 2.49 $\rm \upmu J$, 4.98 $\rm \upmu J$, 7.47 $\rm \upmu J$, 9.96 $\rm \upmu J$, corresponding to P1 to P7. (b) Normalized PL spectra of P1, P4 and P6. Defects from different parameters show similar spectra, including the other spectra that are shown in Figure S4-2(a) in Supporting Information. The proportion of packets at around 620 nm and 720 nm is different with the spectrum in Figure 2, probably because the ensemble is not large enough and the proportion of various defects have not yet become statistically uniform. More information can be found in Section Discussion and Figure S4-2(b) in Supporting Information. (c) The PL image of an spin defect array generated by single pulse energy of 124.8 nJ on quartz substrate. The power of 532-nm pump laser is 2 mW. (d) The distribution of spot size of spin defect array in Figure 3(c). Most of the spot sizes range from 1.7-1.8 $\rm\upmu m$. The yellow dashed line refers to the average size at 1.73 $\rm\upmu m$.}
    \label{Figure 4}
\end{figure*}

The effects of thickness of hBN flakes and energy of single-pulse femtosecond on generated spin defects have been further studied.
Controlling the femtosecond laser parameters unchanged (4.98 $\upmu$J per pulse, 10 pulses), we performed laser irradition on several hBN flakes of different thicknesses on Si/SiO$_2$ substrates.
For hBN flakes of medium thickness (hundred nanometers), the femtosecond laser not only shattered the hBN, but also shattered the substrate (Figure 1(c). Although a small part was dislodged to other locations, the resulting hBN fragments remained mostly around the laser ablation hole, and the PL of spin defects was also mainly concentrated around the hole.
For thinner hBN flakes (less than 100 nm), the results of laser irradiation were generally similar, but the generated hBN fragments mostly fly to other locations (Figure S2(a,b,c) in Supporting Information). We can still detect the PL of spin defects from the fragments left around the irradiation center.
When the thickness of hBN exceeds 1 $\upmu$m, the femtosecond laser is almost completely absorbed by hBN, leaving an inconspicuous hole at the irradiation point (Figure S2(g,h,i) in Supporting Information). The PL map of the sample have a bright luminous round spot at the irradiation spot and ODMR signals can be detected.
These results show that the thickness of hBN has no obvious effect on the generation of spin defects. However, for the hBN flakes with thin and medium thickness on Si/SiO$_2$ substrates, the damage of laser writing is enormous, leaving the irregular fragments of the sample around the irradiation hole, whose orientation could be random and hard to control. When the thickness reaches the $\upmu$m level, this method can be regarded as in situ, which is helpful for further processing and research.

To study the effect of femtosecond laser energy, we use a hBN flake with thickness range in 2.3 $\rm \upmu m$ and 4.3 $\rm \upmu m$ (Figure S2(g,h,i) in Supporting Information) to make sure all laser energy absorbed by hBN.
Varying the single pulse energy (0.25 $\rm \upmu J$, 0.62 $\rm \upmu J$, 1.25 $\rm \upmu J$, 2.49 $\rm \upmu J$, 4.98 $\rm \upmu J$, 7.47 $\rm \upmu J$, and 9.96 $\rm \upmu J$) and fixing the pulse number at 10, we sequentially irradiated 7 spots on the same hBN sample.
As shown in Figure 4(a), more concentrated ensemble of spin defects can be observed in 7 irradiated regions (labeled P1-P7), and the defect density obviously increases as the single pulse energy increases. Especially, the spot P7 reaches a emission rate over $7 \times 10^6$ per second under 1 mW excitation-laser power.
ODMR spectra of all these spots have been detected after annealing.
Figure 4(b) shows the spectra of three spots (P1, P4, P6), and the spectra of the remaining four spots are shown in Figure S4-2(a) in the Supporting Information. These spectra all contain the similar shape, with a packet centered at around 620 nm, indicating that the types and proportion of the defects generated are roughly the same.

We also perform the laser direct writing on hBN flakes on Au and quartz substrates, and annealed at the same conditions. Spin defects with similar properties have been generated on both substrates (PL images in Figure S3(e,f), ODMR spectra in Figure S5-4 and S5-2). The results indicate that substrates seem not to be involved in the process of defect formation. After annealing at 1000 $\rm ^{\circ}$C, Au substrate will bring strong background fluorescence noise. Compared with Si/SiO$_2$ substrate, the damage of quartz substrate is small under the same femtosecond laser energy, which is due to the transparency of quartz. Therefore, quartz substrate is a good choice for generating spin defects.

To generate spin defects with minimal damage to hBN which is important for potential applications, we further optimized the preparation conditions. By choosing quartz substrate, we successfully written an array of emitters with yield of almost 100\% with single laser pulse energy of 124.8 nJ upon a 556 nm hBN flake. The PL image is shown in Figure 4(c), and spectra of three spots are shown in Figure S4(g-i) in Supporting Information. We have performed ODMR measurements on over half of the spots. Their ODMR spectra all have been observed and three of them are shown in Figure S5-2 in Supporting Information. According to the AFM image shown in Figure S2(k) in the Supporting Information, the average spot size is 1.73 $\rm\upmu m$, indicating minor damage to the sample. The statistical estimates of spot sizes are shown in Figure 4(d).

\section*{Discussions}
The experimental results show that both laser irradiation and annealing are indispensable factors in the generation of the bright defect ensemble. We speculate that the high-energy femtosecond laser pulse breaks the covalent bonds in hBN, and the subsequent vacuum high-temperature annealing causes the atoms to reorganize to form new stable chemical bonds, resulting in fluorescent spin defects.
The spectra of defect ensemble are all basically a large packet with two obvious peaks near 620 nm and 720 nm, but the relative proportions of these two peaks in the spectra of defect ensemble generated on different hBN flakes are different. ODMR signals were detected experimentally in the defect ensembles that we generated using this method, and by using 700-nm long-pass (FELH0700, Thorlabs) and short-pass (FESH0700, Thorlabs) filters, positive ODMR spectra for these two different spectral peaks can be detected, respectively. We speculate that this method produces a variety of different defects and the further cryogenic spectra of these defect ensembles give corresponding evidence (Figure S6-1 in Supporting Information). This can be attributed to the complex process of high-energy laser irradiation and chemical bond recombination, in which not only native defects such as vacancies, antisite defects and interstitial defects are generated, but also the doping defects formed by the original atoms in hBN flakes \cite{SolventEx} or even laser-ionized atoms in air. At the mean time, the varying degrees of damage by laser writing, the uneven distribution of impurities in hBN flakes and the statistical nonuniformity of the small defect ensemble, jointly contribute to the different proportion of spectral peaks (Figure S4-2(b) in Supporting Information). Further study of defect species is beyond the scope of this paper and requires more research in the future.

\section*{Conclusion}
In conclusion, we propose a deterministic technique to generate spin defects in nanoscale hBN flakes by femtosecond laser irradiation. Femtosecond laser breaks the original lattice and subsequent high-temperature vacuum annealing restructure the lattice to create specific defects. In the presence of a magnetic field perpendicular to the substrate, we observed positive single-peak ODMR signals.
Further results show that the thickness of hBN flakes have little effect on the generated defect species, but the number of defects and the damage exerted on the sample is positively correlated with pulse energy. The generation process can be further optimized by reducing the pulse energy and applying quartz substrate for minimum damage to sample, which will be helpful for future research on solid-state spin centers in hBN.
Our work provides a new simple tool to engineer spin defects and will motivate more endeavors for researches and applications of spin-based technologies, such as nanoscale magnetometer sensor array. In addition, further theoretical research are needed to give insight into the structure and intrinsic nature of these defects.

\section*{Method}

\textbf{Sample preparation.}
The hBN flakes purchased from HQ Graphene were mechanical exfoliated and transferred onto the Si substrate with 285-nm $\rm SiO_2$ top layer. The substrates are coated with Au marks as the location coordinates in the microscope, and cleaned by acetone, isopropyl alcohol and deionized water successively before using. The same methods are used to fabricated Au film with thickness of 210 nm on Si/SiO$_2$ substrate. The quartz substrate is commercially purchased with thickness of 0.7 mm. Laser irradiation was done using femtosecond laser micromachining \cite{Memory1, Memory2} from WOPhotonics (Altechna RD Ltd, Lithuania) with a 210-fs pulse duration and 1 kHz repetition rate. An 100$\rm\times$ objective with numerical aperture (N.A.) = 0.7 was used to focus the laser on the sample.
After 10-pulse irradiation with energy varying from 124.8 nJ to 9.96 $\rm \upmu J$, the samples were annealed in the tube furnace in high vacuum environment ($\rm\textless$ $\rm 10^{-6}$ torr) with temperature of 1000 $\rm ^{\circ}C$ for 2 hours.

\textbf{Experimental setup.}

Atomic force microscope from Bruker (Dimension Icon) was used for thickness characterization of the hBN flakes. The AFM probe used in tapping mode is RTESP-300 from Bruker.

A 532-nm excitation laser with bandwidth less than 3 nm was transmitted through a fused silica plate beamsplitter (BSW26R, Thorlabs) and focused on the samples through a N.A. = 0.9 objective (Olympus MPLFLN100xBDP). The fluorescence was collected by the same objective, reflected by the same beamsplitter, filtered by a 550-nm long-pass filter (FELH0550, Thorlabs), and finally guided through a 9-$\upmu$m-core-diameter fiber to a silicon-based single photon avalanche diode (SAPD) for detection.
In the microwave system, a silver wire with a diameter of 9 $\rm \upmu m$ was suspended in close proximity to the laser ablation hole to deliver the microwave field generated by a synthesized signal generator (SSG-6000RC, Mini-Circuits) and amplified by a power amplifier (ZHL-20W-13SW+, Mini-Circuits). An electric magnet is placed under the sample to provide magnetic field perpendicular to the substrate surface.

\section*{Acknowledgments}

This work is supported by the Innovation Program for Quantum Science and Technology (No. 2021ZD0301200), the National Natural Science Foundation of China (Nos. 12174370, 12174376, and 11821404), the Youth Innovation Promotion Association of Chinese Academy of Sciences (No. 2017492), the Open Research Projects of Zhejiang Lab (No.2021MB0AB02), the Fok Ying-Tong Education Foundation (No. 171007). This work was partially carried out at the USTC Center for Micro and Nanoscale Research and Fabrication.

\onecolumngrid
\clearpage
\begin{center}
  \textbf{\ Supporting Information}\\[.2cm]
\end{center}
\section*{S1. Supplementary notes for defect information}
\begin{figure*}[htbp]
    \centering
    \includegraphics[width=0.8\linewidth]{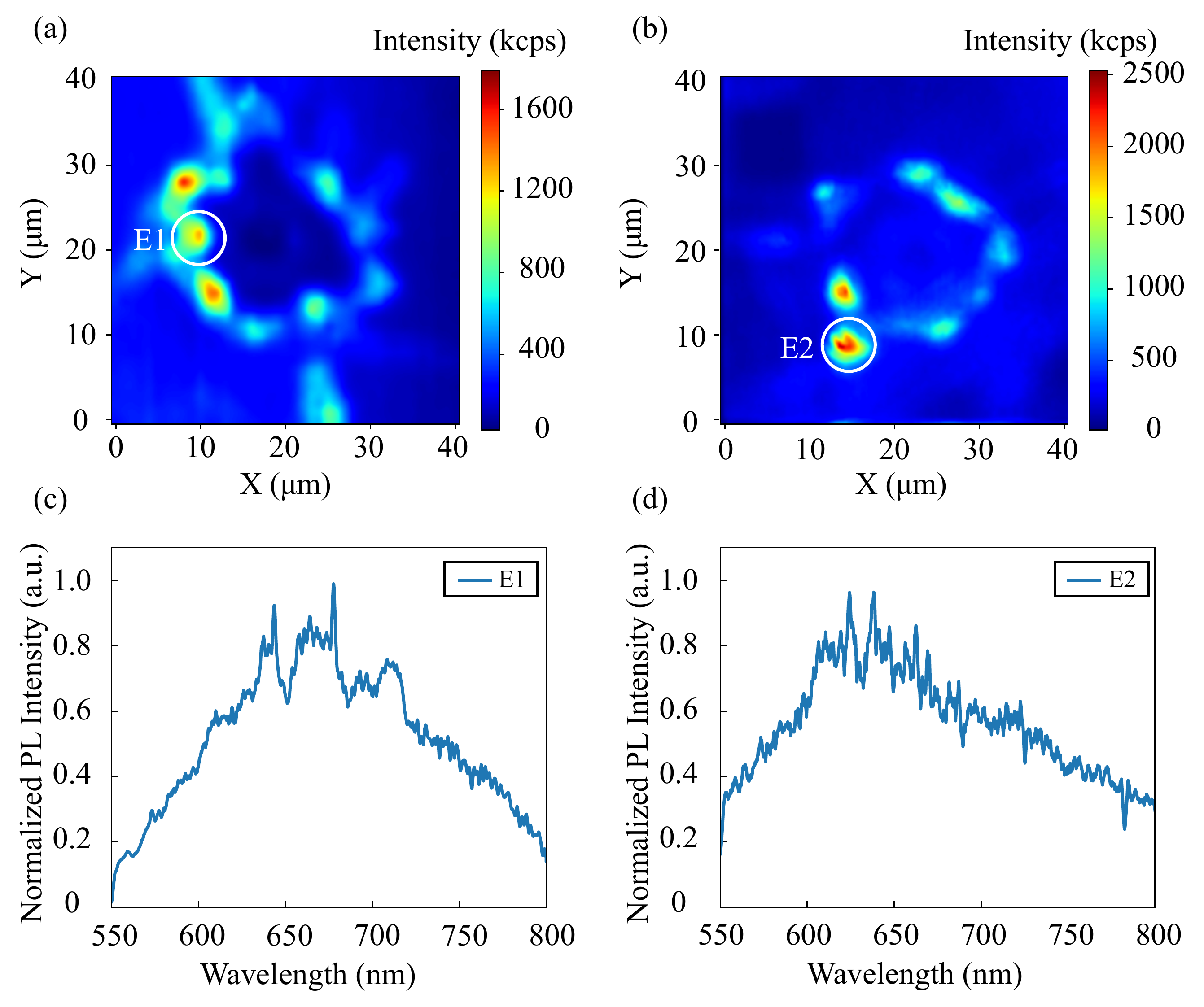}
    \renewcommand{\thefigure}{S1}
    \caption{(a) White circled defect E1 refers to the defect in Figure 3(a,b). (b) Defect E2, corresponds to the defect in Figure 3(c,d). (c,d) Room temperature spectra of E1 and E2, respectively.}
    \label{Figure_S1}
\end{figure*}
Defects E1 and E2 correspond to Figure 3(a,b) and Figure 3(c,d) in main text, respectively. Figure S1(c) and Figure S1(d) show their spectra at room temperature. Emissions of the two defects are collected using a coupling lens coated with a high transmittance film at 650 - 1050 nm. The excitation power is 1 mW with wavelength of 532 nm.
\newpage

\section*{S2. Morphological characterization of laser-irradiated hBN flakes with different thickness}
\begin{figure*}[htbp]
    \centering
    \includegraphics[width=0.95\linewidth]{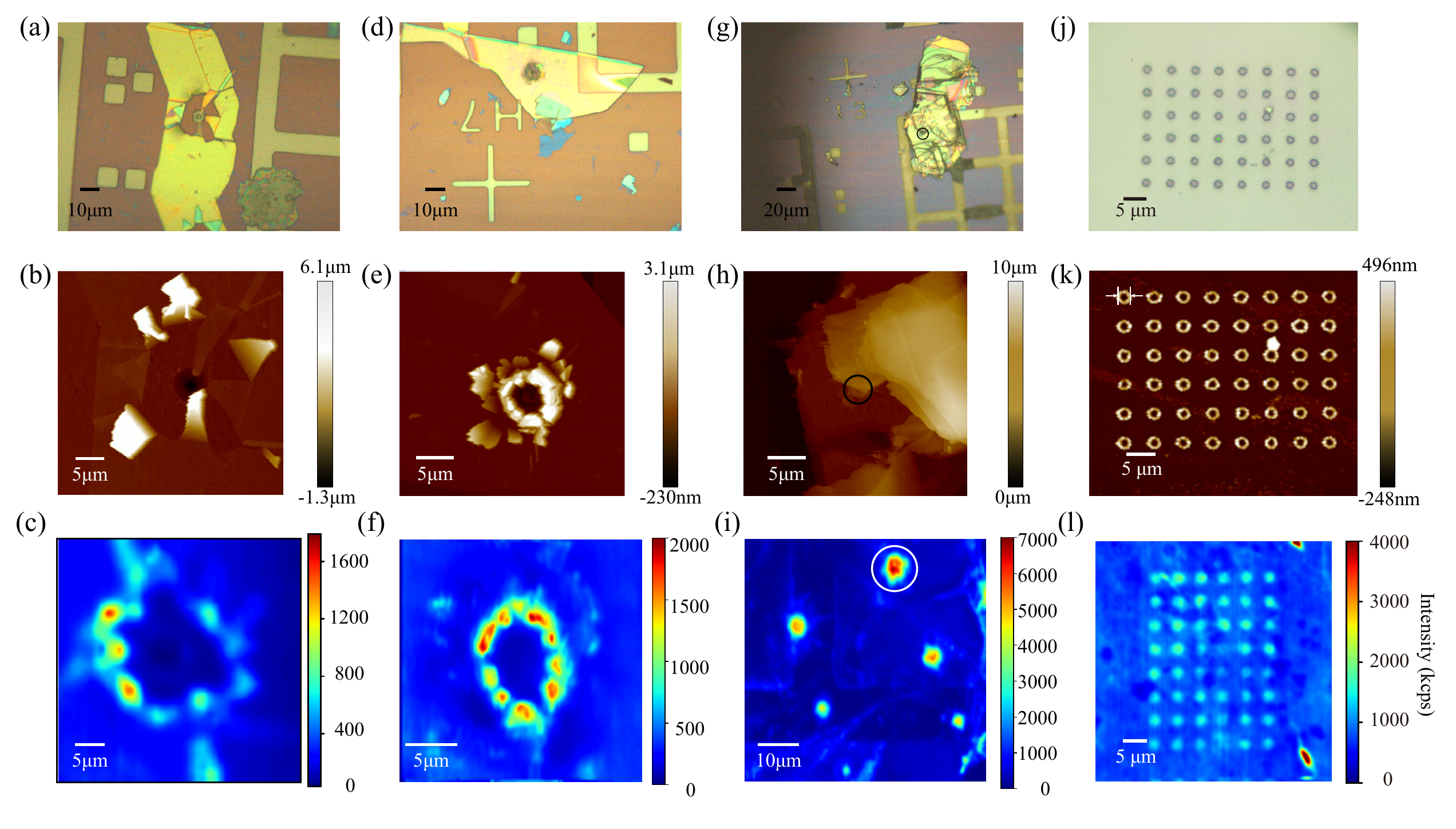}
    \renewcommand{\thefigure}{S2}
    \caption{(a,d,g,j) Optical images of hBN samples. The first three samples are on Si/SiO$_2$ substrates, and the last one is on quartz substrate. (b,e,h,k) Corresponding AFM characterized images. The thickness of the samples are 80 nm, 255 nm, above 2 $\rm \upmu$m, and 556 nm from left to right, respectively. The diameters of the laser written regions of the first two samples are 20 $\rm \upmu$m and 5 $\rm \upmu$m, respectively. The damage by laser irradiation of sample in (g) is not clear enough to distinguish. The last sample in (k) shows the average of spot size is 1.73 $\rm\upmu m$ and the maximum depth is 248 nm. (c,f,i,l) PL images of the corresponding samples after annealing.}
    \label{Figure_S2}
\end{figure*}

Characterization of samples with different thickness under optical microscope, AFM and PL confocal images are shown in Figure S2. The first three samples are on Si/SiO$_2$ substrates, and the last one is on quartz substrate. The thickness of hBN flakes are 80 nm, 255 nm, above 2 $\rm \upmu m$, and 556 nm from left to right, respectively. The diameters of the holes on first two samples caused by laser writing are 20 $\rm \upmu m$ and 5 $\rm \upmu m$, and the third is too small to distinguish. The results show that with the parameters of the laser irradiation fixed, the thicker the sample, the smaller the holes left, which indicates the better absorption of femtosecond laser power and less energy loss. The last sample written with optimized parameters (124.8 nJ per pulse and quartz substrate) presents holes of average spot size of 1.73 $\rm\upmu m$ and the maximum depth of the sample is 248 nm.
\newpage

\section*{S3. Photoluminescence Maps Before and After Annealing}
\begin{figure*}[htbp]
    \centering
    \includegraphics[width=0.95\linewidth]{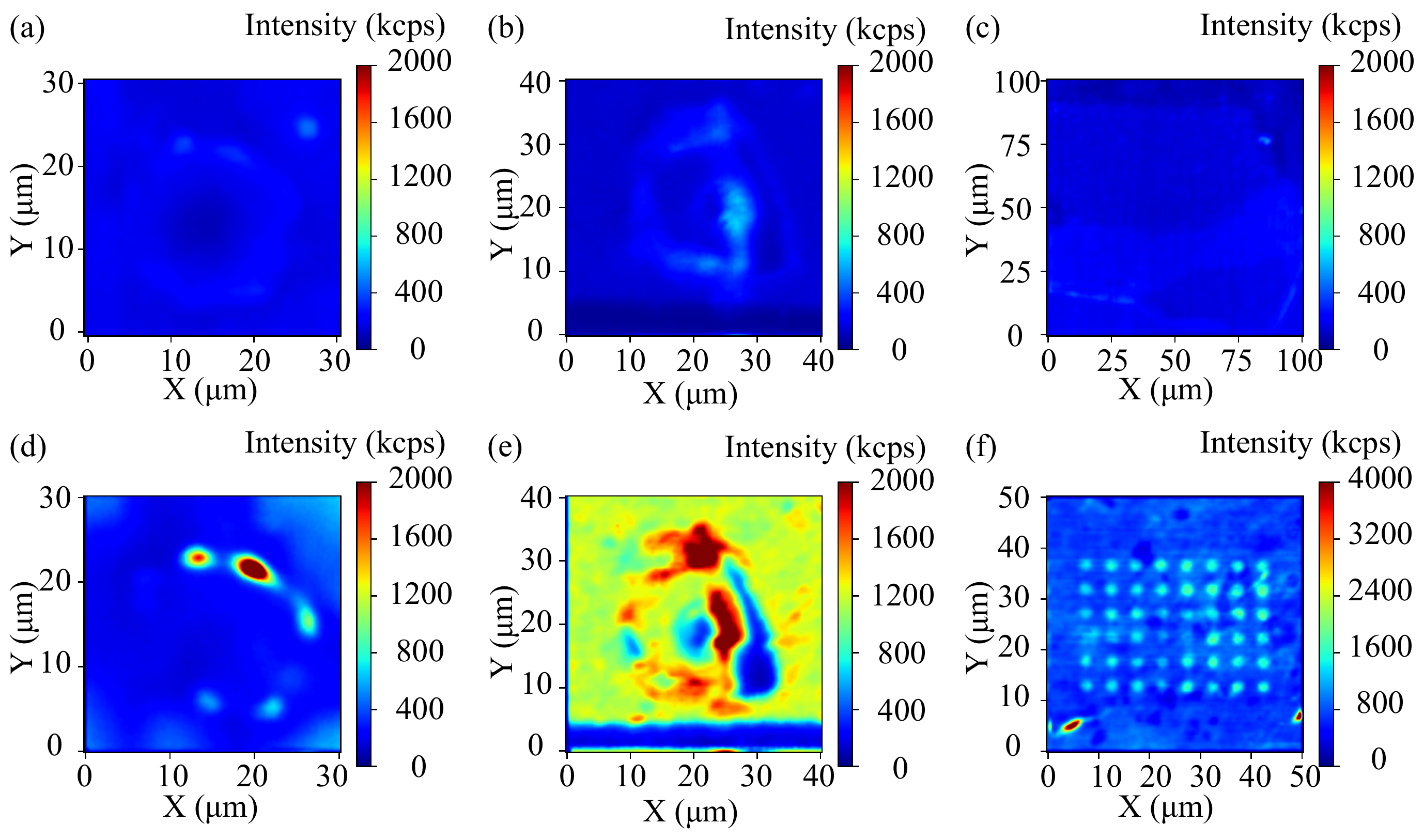}
    \renewcommand{\thefigure}{S3}
    \caption{(a) The PL map of sample on Si/SiO$_2$ substrate after femtosecond laser irradiation (4.98 $\rm\upmu J$) and before annealing. (b) The PL map of sample on Au substrate after femtosecond laser irradiation (4.98 $\rm\upmu J$) and before annealing. (c) The PL map of sample on quartz substrate after femtosecond laser irradiation (124.8 nJ)  and before annealing. (d-f) The corresponding PL maps after annealing. The defects become much brighter after annealing.}
    \label{Figure_S3}
\end{figure*}

Figure S3 shows the corresponding PL maps before (a-c) and after (d-f) annealing. Annealing makes the irradiated spots much brighter, playing a key role in activating the defects.

\newpage

\section*{S4. Spectra of different Samples}
\begin{figure*}[htbp]
    \centering
    \includegraphics[width=0.95\linewidth]{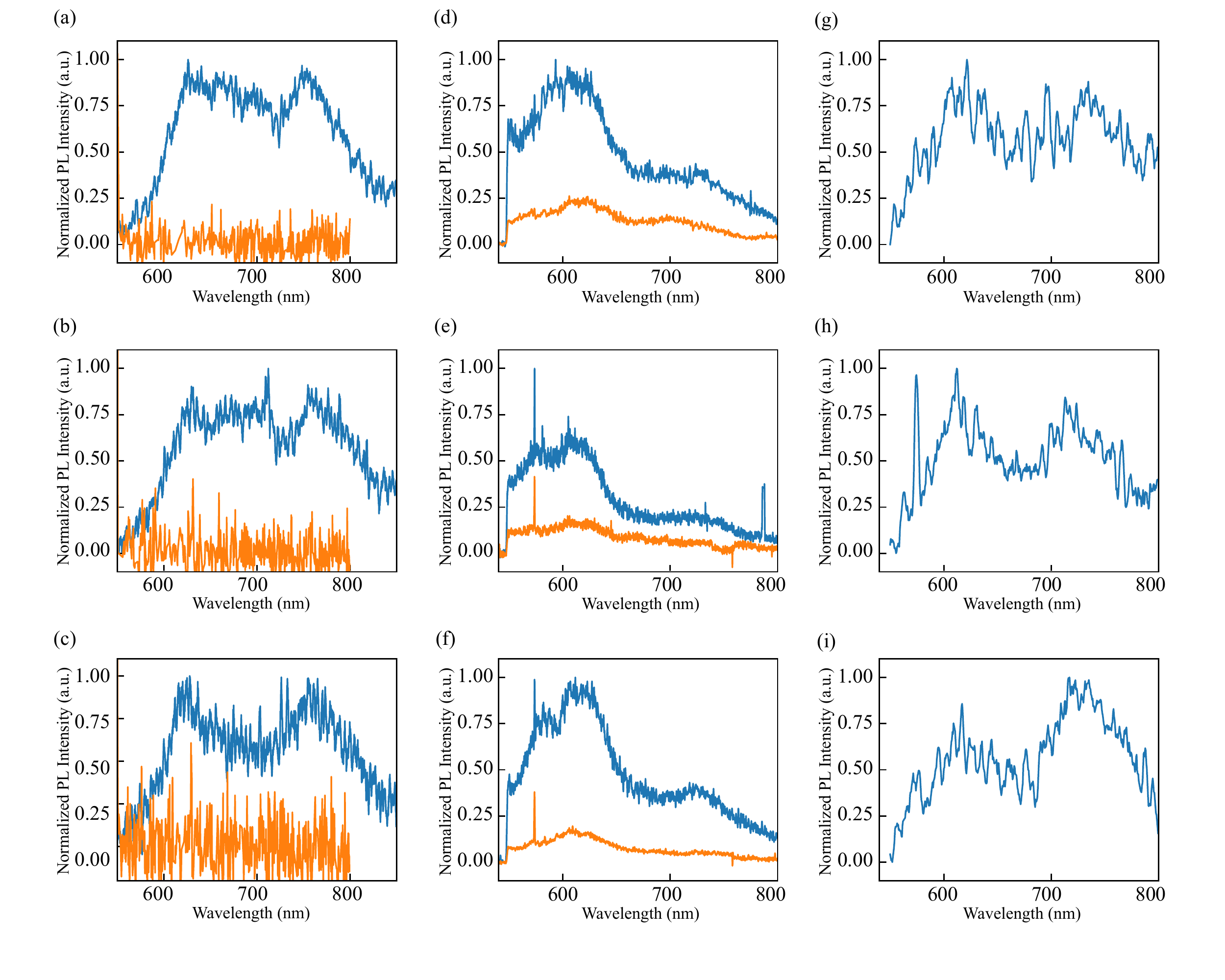}
    \renewcommand{\thefigure}{S4-1}
    \caption{(a-c) The spectra of irradiated samples on $\rm Si/SiO_2$ substrates before and after annealing with single pulse energy of 4.98 $\rm\upmu J$. (d-f) The spectra of samples on Au substrates with single pulse energy of 4.98 $\rm\upmu J$ before and after annealing. (g-i) The spectra of irradiated sample on quartz substrate generated with single pulse energy of 124.8 nJ after annealing.}
    \label{Figure_S4-1}
\end{figure*}

Spectra of different samples are shown in Figure S4-1. In (a-c), three spots of emitters were generated on $\rm Si/SiO_2$ substrate with single pulse energy of 4.98 $\rm\upmu J$ and their spectra before (yellow lines) and after (blue lines) annealing are shown. The shape of spectra before annealing is difficult to recognize due to the low emission rates. Three spots of emitters were generated on Au with single pulse energy of 4.98 $\rm\upmu J$ and their spectra before (yellow lines) and after (blue lines) annealing are also shown in (d-f). As the irradiated spots cannot be recognized before annealing in Figure S3(c), only the spectra of randomly selected emitters after annealing are shown in Figure S4-1(g-i).

\newpage
\begin{figure*}[htbp]
    \centering
    \includegraphics[width=0.95\linewidth]{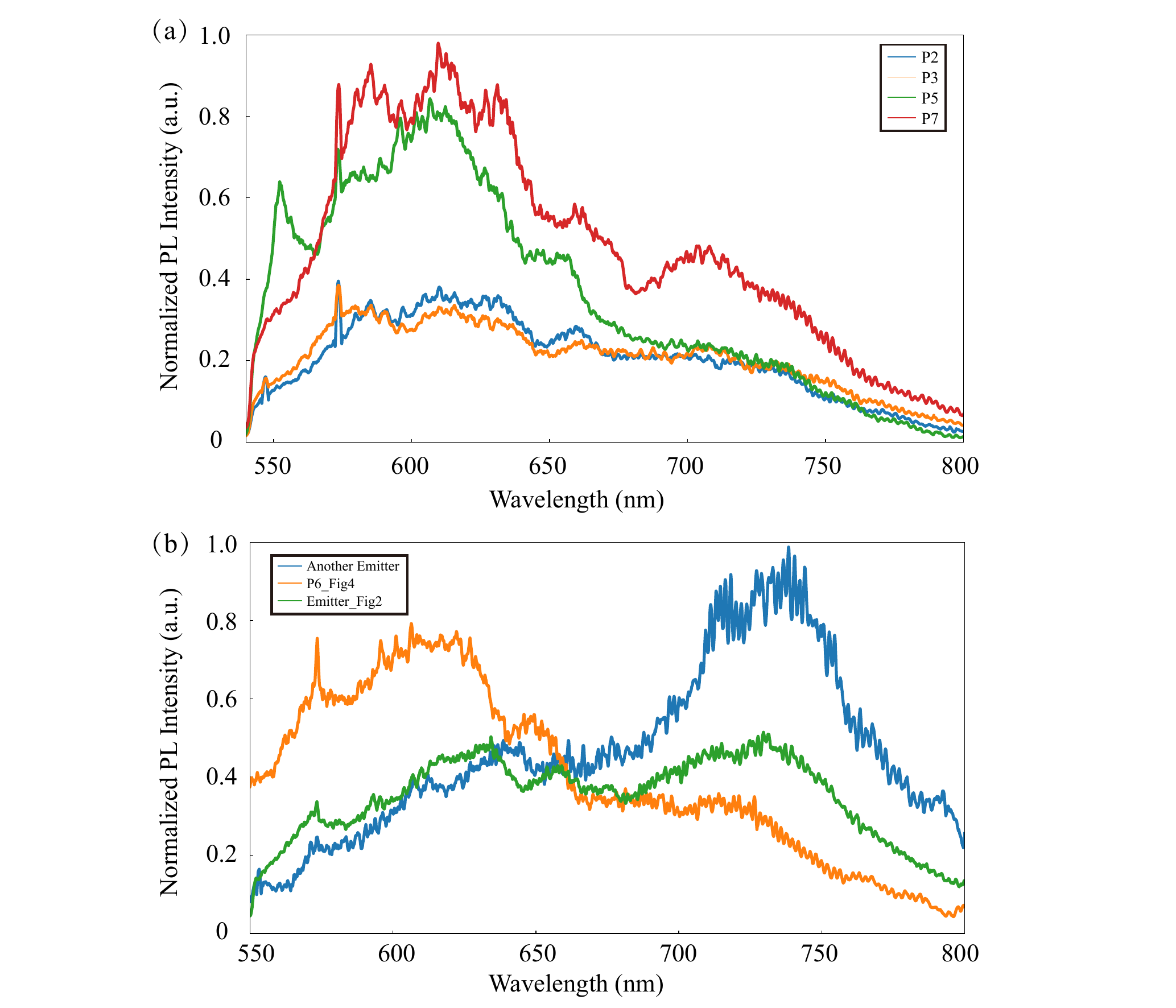}
    \renewcommand{\thefigure}{S4-2}
    \caption{(a) The other four spectra as a complement for Figure 4 in the main text. (b) Several representative spectra. The blue line shows a dominant packet at around 720 nm, different from the spectra in Figure 2 and Figure 4.}
    \label{Figure_S4-2}
\end{figure*}

The rest four spectra of the defects (in Figure 4(a)) with different femtosecond laser writing energy (P2, P3, P5, P7) are shown in Figure S4-2(a) and the peak at 573 nm refers to the Raman peak of hBN. The count rate is positively correlated with the femtosecond direct writing energy. These defects exhibit similar spectra, indicating that they are the same kind of defects with similar proportion. Figure S4-2(b) presents some typical spectra showing different dominant peaks. Components of spectrum packets at around 620 nm and 720 nm occupy different proportions. The spectrum shown by the solid blue line displays a packet dominated by 720 nm, P6 in Figure 4 (orange line) is dominated by 620 nm, and the emitter in Figure 2 (green line) having a similar proportion of the two peaks. The difference may come from the nonuniformity of various defects in the small ensemble during the generation process. ODMR signals can be detected in all these samples.

\newpage

\section*{S5. Supplementary ODMR Measurements}
\begin{figure*}[ht]
    \centering
    \includegraphics[width=1\linewidth]{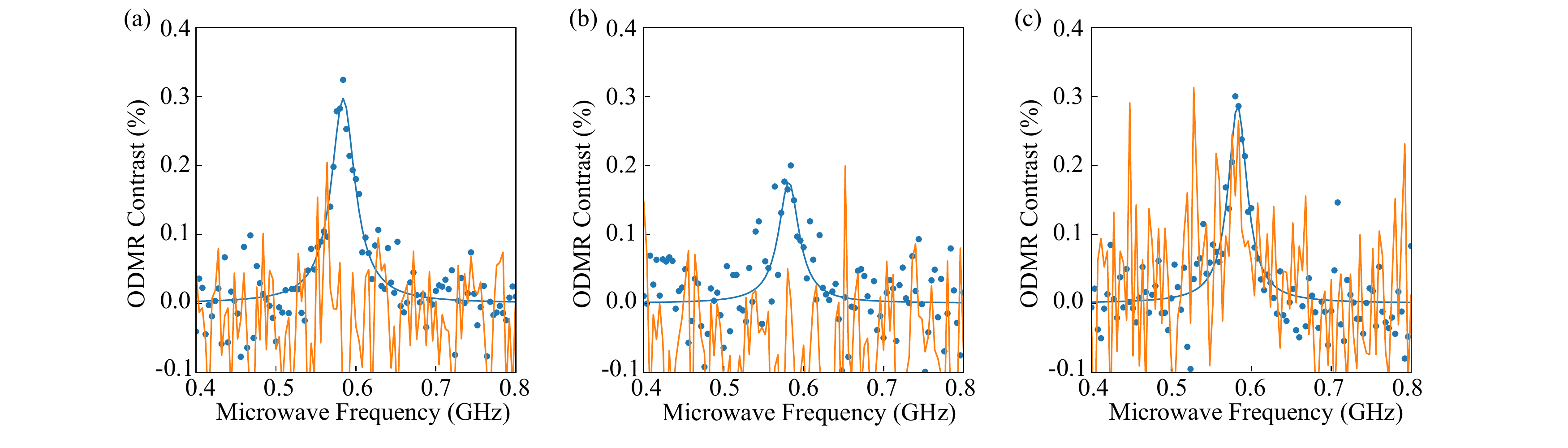}
    \renewcommand{\thefigure}{S5-1}
    \caption{ODMR spectra before (yellow lines) and after (blue lines and dots) annealing of the irradiated samples on Si/SiO$_2$ substrate. Before annealing, no obvious signals can be detected, while obvious ODMR signals can be observed after annealing.}
    \label{Figure_S5-1}
\end{figure*}

Figure S5-1 shows the ODMR spectra before (yellow lines) and after (blue lines and dots) annealing of sample in Figure S3(a,d) under 20.1 mT. As the figure presents, the ODMR signals are obvious after annealing while no obvious signals can be detected before annealing, proving that annealing is a necessary step to activate the spin defects.

\begin{figure*}[ht]
    \centering
    \includegraphics[width=1\linewidth]{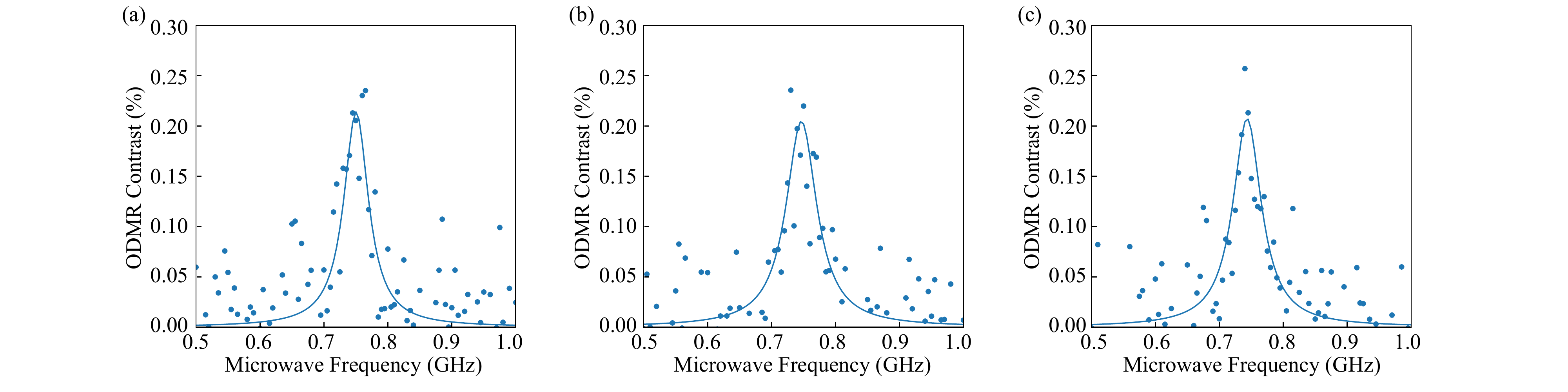}
    \renewcommand{\thefigure}{S5-2}
    \caption{ODMR spectra of irradiated samples on quartz substrate.}
    \label{Figure_S5-2}
\end{figure*}

\begin{figure*}[ht]
    \centering
    \includegraphics[width=1\linewidth]{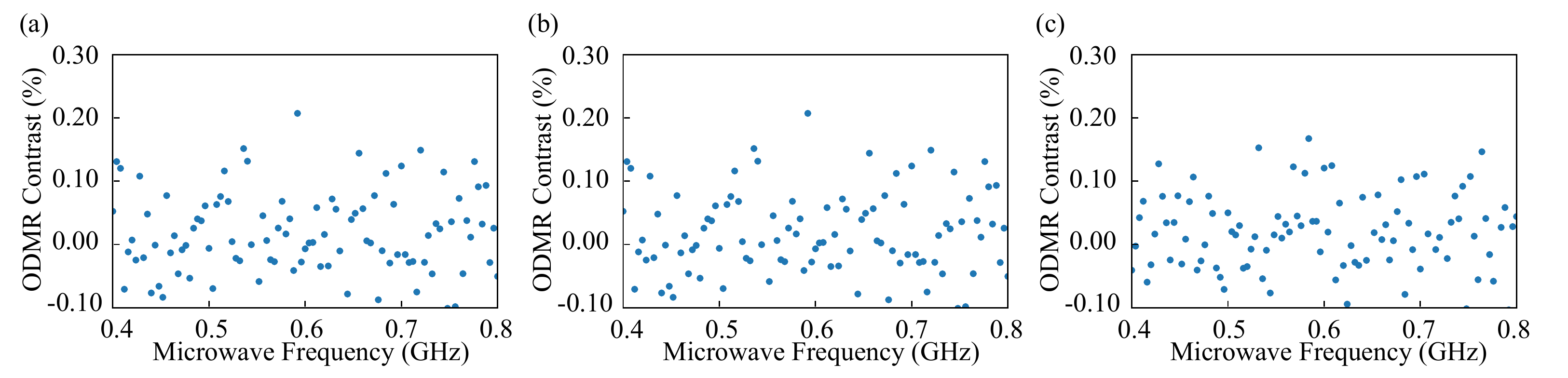}
    \renewcommand{\thefigure}{S5-3}
    \caption{ODMR spectra of irradiated samples on quartz substrate with external magnetic field parallel to the substrate surface. No obvious signals have been observed.}
    \label{Figure_S5-3}
\end{figure*}

ODMR signals shown in Figure S5-2 are from the randomly selected spots in the emitter array of Figure 4(c). Similar ODMR signals can be detected under 26 mT. When the magnetic field direction is changed to be parallel to the substrate, no obvious ODMR signals have been observed (Figure S5-3) under 20 mT.

\begin{figure*}[ht]
    \centering
    \includegraphics[width=1\linewidth]{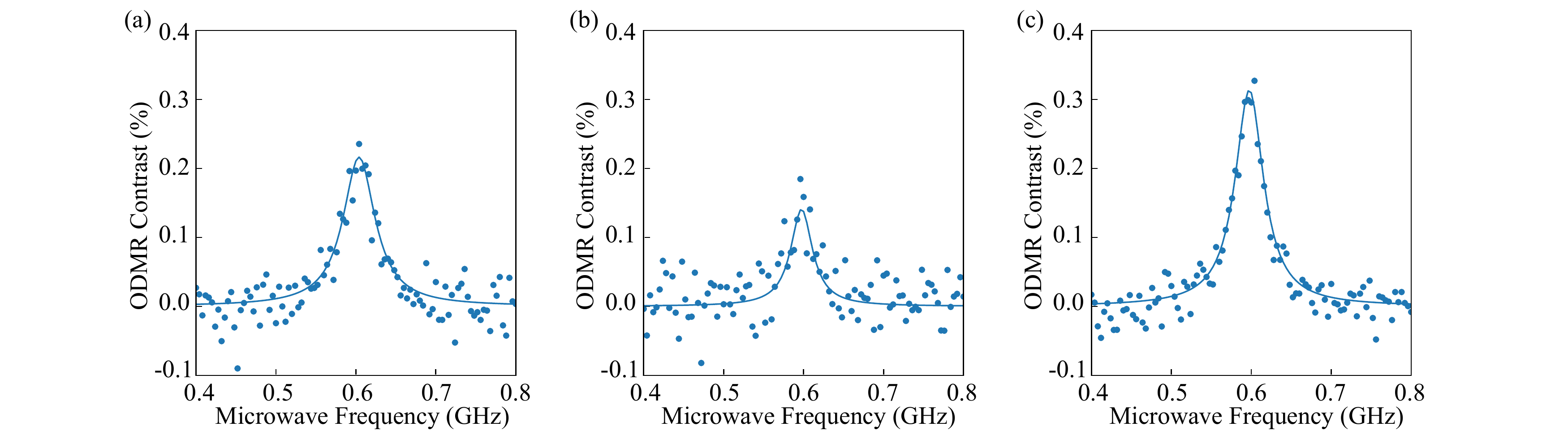}
    \renewcommand{\thefigure}{S5-4}
    \caption{ODMR spectra of irradiated samples on Au substrate.}
    \label{Figure_S5-4}
\end{figure*}

The ODMR spectra of irradiated samples on Au substrate under 21.1 mT, corresponding to the sample in Figure S3(b,e) are shown in Figure S5-4.

\newpage

\section*{S6. Cryogenic spectra of different samples}
\begin{figure*}[ht]
    \centering
    \includegraphics[width=0.95\linewidth]{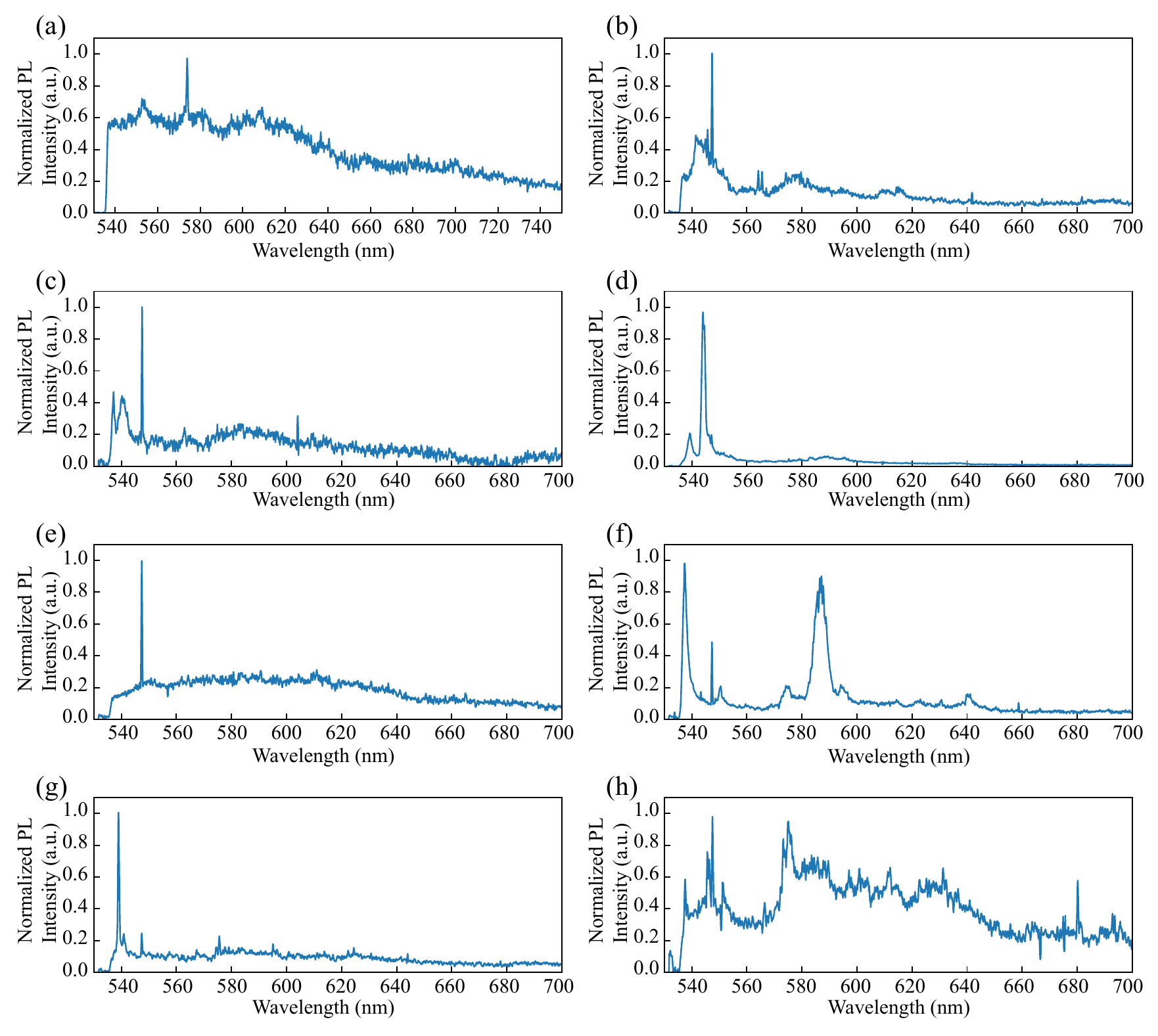}
    \renewcommand{\thefigure}{S6-1}
    \caption{(a) The spectrum at 4 K corresponding to the defect in Figure 2, still manifesting as a large packet. (b to h) Other cryogenic spectra of femtosecond laser written samples, showing relatively narrow peaks at around 538 nm, 540 nm, 545 nm and 585 nm. The narrow peaks at 547 nm refer to the Raman peak of $\rm SiO_2$.}
    \label{Figure_S6-1}
\end{figure*}
\newpage
\begin{figure*}[ht]
    \centering
    \includegraphics[width=1\linewidth]{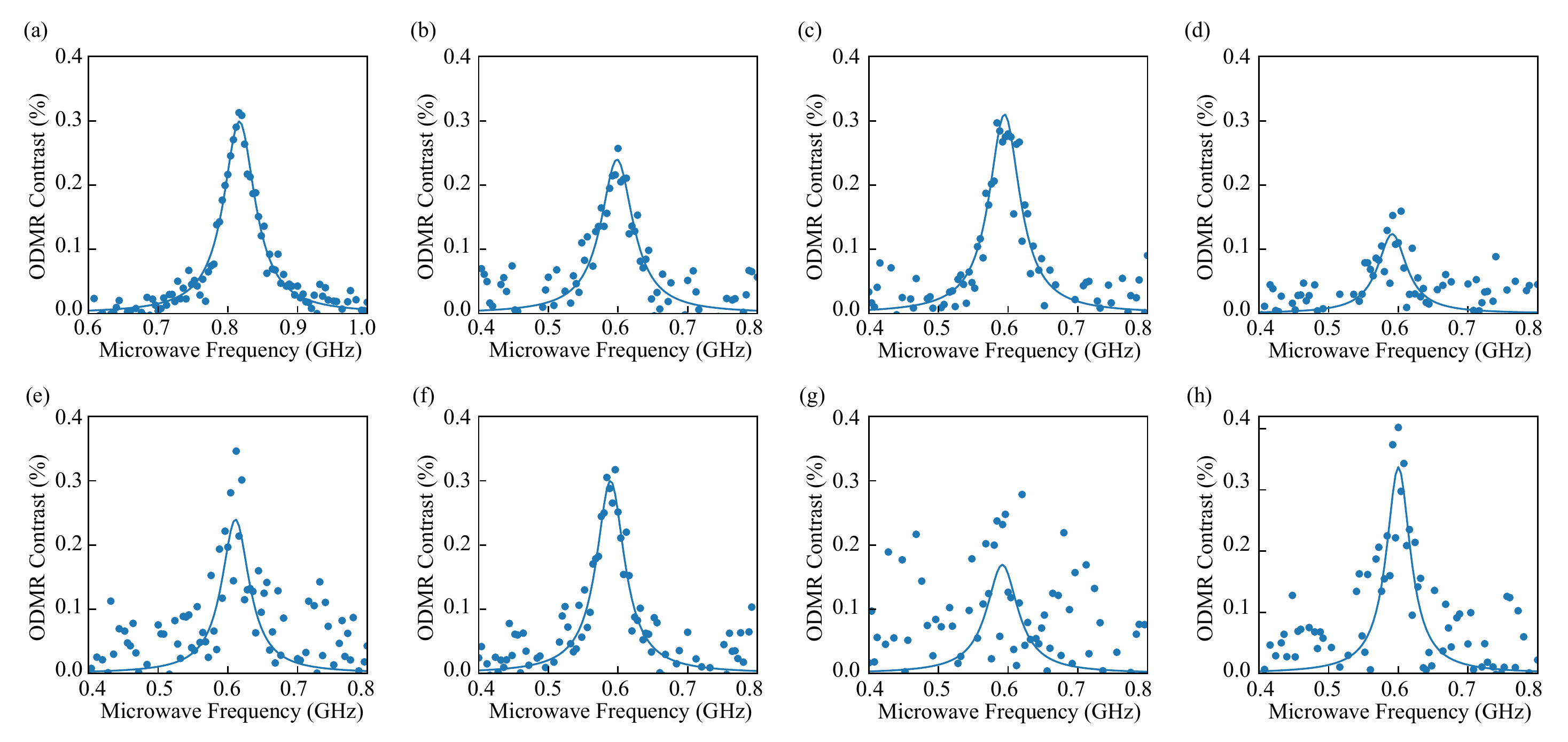}
    \renewcommand{\thefigure}{S6-2}
    \caption{ODMR spectra of different samples corresponding to that in Figure S6-1.}
    \label{Figure_S6-2}
\end{figure*}

Different cryogenic spectra at 4 K are shown in Figure S6-1, which were all generated under the same femtosecond laser condition: 4.98 $\rm\upmu J$ single pulse energy and 10 pulses. Spectrum in Figure S6-1(a) shows the cryogenic spectrum of the defect in Figure 2, whose zero phonon line cannot be distinguished under cryogenic condition. Figure S6-1(b-f) are on a same sample and (g,h) are on another same sample. Some narrow peaks at around 540 nm (Figure S5(c,d,f,g)), 545 nm (Figure S5(d)) and 585 nm (Figure S5(f)) can be observed while some spectra are also broad packets (Figure S5(b,e,h)). The results of cryogenic spectra show that various defects are generated. ODMR signals are also observed in each defect, as shown in Figure S6-2. Figure S6-2(a) was detected under the magnetic field of 29.5 mT, while Figure S6-2(b-h) are detected under 21.1 mT.
\newpage

\section*{S7. Measurement of Single Photon Emitters}
\begin{figure*}[ht]
    \centering
    \includegraphics[width=0.95\linewidth]{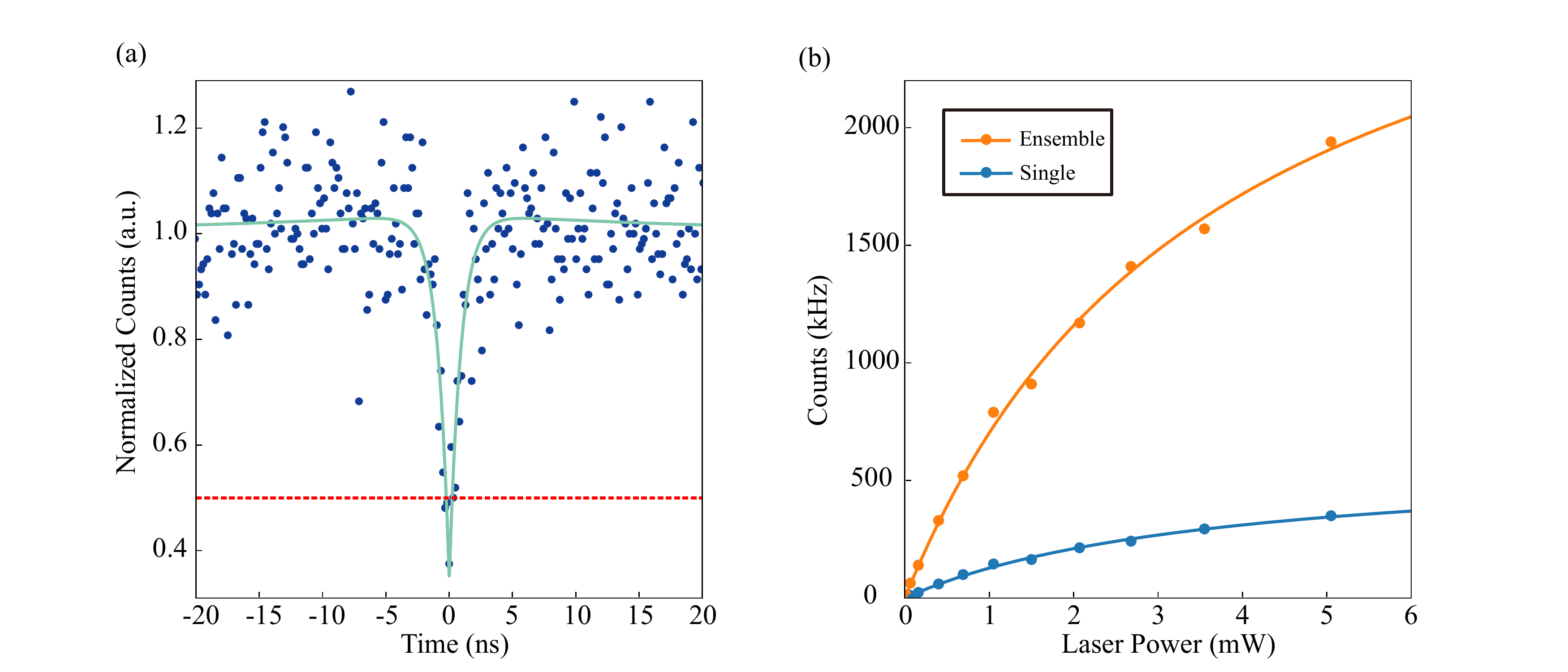}
    \renewcommand{\thefigure}{S7}
    \caption{(a) The antibunching second-order time correlation $g^{(2)}(\tau)$ of the defect (Figure S6-1(a)). The fitting result using three-level model shows $g^{(2)}(0)$=0.34. (b) The fluorescence intensity of defects plotted vs laser power. The blue line refers to the single photon emitter in Figure S7(a) while the yellow line refers to a typical defect ensemble, showing saturation laser power at 3.67 mW and 3.72 mW, respectively.}
\end{figure*}

Second order time correlation measurement was performed on the emitter in Figure S6-1(d), shown in Figure S7(a).
The green fitted line presents $g^{(2)}(0)=0.34$ under 0.45 mW 532-nm laser pumping, and the red dashed line at 0.5 represents the boundary of a single photon emitter. The antibunching second-order time correlation $g^{(2)}(\tau)$ is fitted with three-level model as function below:
$$G^{(2)}(\tau-\tau_0)=N\{1-\rho^2[(1+a)e^{-|\tau-\tau_0|/t_1}-ae^{-|\tau-\tau_0|/t_2}]\},$$
and $g^{(2)}(\tau-\tau_0)=G^{(2)}(\tau-\tau_0)/N$. $t_1$ and $t_2$ refer to the lifetime of excited state and metastable state, respectively. $\tau_0$, $\rho$, $a$ and $N$ represent constant delay time, single-photon purity, population weight and normalized coefficient, respectively.

Figure S7(b) show the saturation emission behavior of defects under 532-nm laser. The blue line is the single photon emitter (Figure S6-1(a)) and the the yellow refers to a typical defect ensemble, and they saturate at 3.67 mW and 3.72 mW, respectively. By applying 532 nm laser excitation with power of 3.72 mW, the count rates of the single emitter and the typical ensemble are 0.60 MHz and 3.32 MHz, respectively. At the same time, due to the increase of background noise, $g^{(2)}(0)$ of the single defect increases to 0.70. The calibrated count rate can be derived by $I_{Single}=\sqrt{1-g^2(0)}I_{Total}$, where $I_{Total}$ refers to the total count: 0.60 MHz. The calculated calibrated count rate of single defect is 0.33 MHz. As the count rate of the typical ensemble is 3.32 MHz, we can roughly estimate that there are 10 defects per spot.

\newpage
\section*{S8. Raman Characterization}
\begin{figure*}[ht]
    \centering
    \includegraphics[width=1\linewidth]{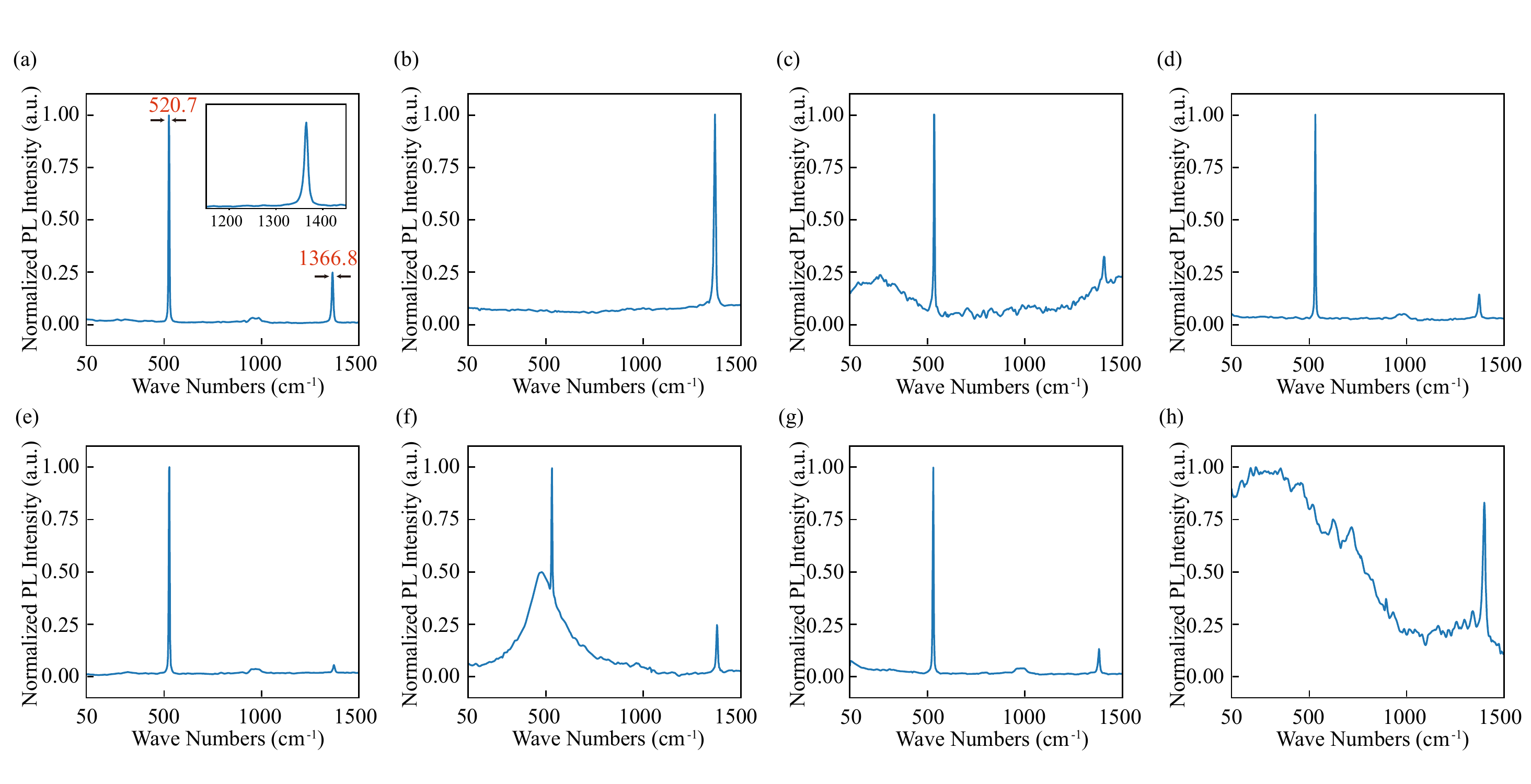}
    \renewcommand{\thefigure}{S8}
    \caption{Raman spectra of different samples. The samples corresponding to (a,c-g) are on Si/SiO$_2$ substrates, the sample corresponding to (b) is on quartz substrate, and the sample corresponding to (h) is on Au substrate. The peak at 520.7 $\rm cm^{-1}$ refers to the Raman peak of Si, and the peak at 1366.8 $\rm cm^{-1}$ refers to the Raman peak of hBN. The small box in (a) indicates no existance of Raman peak of cubic boron nitride ($\rm 1295$ $\rm cm^{-1}$).}
    \label{Figure_S8}
\end{figure*}

Figure S8 shows the Raman spectra of all the typical presented samples. The Raman spectra of Figure S8(a-h) refer to defects in Figure 4(a), Figure 4(c), Figure 2, Figure S1(a), Figure S1(b), Figure S6-1(d), Figure S3(a,d) and Figure S3(b,e). The Raman peak of 520.7 $\rm cm^{-1}$ refers to Si, which can not be detected in Figure S8(b,h), whose substrates are quartz and Au, respectively. The peak at 1366.8 $\rm cm^{-1}$ refers to the Raman peak of hBN, which can be observed in all the samples. The small figure in Figure S8(a) proves no existance of other materials such as cubic boron nitride, whose Raman peak locates at around $\rm 1295$ $\rm cm^{-1}$.

\end{document}